\documentclass[12pt]{article}
\usepackage{graphics}
\usepackage{amssymb}

\if@twoside  m
\else
\fi \topmargin 5mm \headheight 0mm \headsep 0mm \textheight
225truemm \textwidth 150truemm
\newcommand{\QED}{\mbox{\rule[-1.5pt]{6pt}{10pt}}}
\newcommand{\N}{\mathbb{N}}
\newcommand{\R}{\mathbb{R}}
\newcommand{\C}{\mathbb{C}}
\newcommand{\ie}{{\em i.e.}}
\newcommand{\cf}{{\em cf. }}
\newcommand{\rhs}{{r.h.s.\ }}
\newcommand{\lhs}{{l.h.s.\ }}
\newcommand{\BB}{{\cal B}}
\newcommand{\OO}{{\cal O}}
\newtheorem{claim}{Claim}[section]
\newtheorem{theorem}[claim]{Theorem}
\newtheorem{proposition}[claim]{Proposition}
\newtheorem{lemma}[claim]{Lemma}
\newtheorem{remark}[claim]{Remark}
\begin{document}

\title{Anomalous Pauli electron states \\ for magnetic
fields with tails}
\author{P.~Exner,$^{a,b}$ M.~Hirokawa,$^{c}$ and
O.~Ogurisu$^{d}$}
\date{}
\maketitle
\begin{quote}
{\small \em a) Department of Theoretical Physics, Nuclear Physics
Institute, Academy of \phantom{e)x}Sciences, 25068 \v Re\v z,
Czech Republic \\
 b) Doppler Institute, Czech Technical University, B\v{r}ehov{\'a} 7,
11519 Prague, \phantom{e)x}Czech Republic \\
 c) Department of Mathematics, Faculty of Science, Okayama
 University,\\ \phantom{e)x}3-1-1 Tsushima-naka, Okayama 700-8530, Japan \\
 d) Department of Computational Science, Faculty of Science, Kanazawa \\
\phantom{e)x}University, Kakuma--machi Kanazawa, Ishikawa 920,
Japan \\
 \rm \phantom{e)x}exner@ujf.cas.cz, hirokawa@math.okayama-u.ac.jp,
 \\ \phantom{e)x}ogurisu@lagendra.s.kanazawa-u.ac.jp} \vspace{8mm}

\noindent {\small We consider a two-dimensional electron with an
anomalous magnetic moment, $g>2$, interacting with a nonzero
magnetic field $B$ perpendicular to the plane which gives rise to
a flux $F$. Recent results about the discrete spectrum of the
Pauli operator are extended to fields with the
$\OO(r^{-2-\delta})$ decay at infinity: we show that if $|F|$
exceeds an integer $N$, there is at least $N+1$ bound states.
Furthermore, we prove that weakly coupled bound states exist under
mild regularity assumptions also in the zero flux case. }
\end{quote}

%%%%%%%%%%%%%%%%%%%%%%%%%%%%%%%%%%%%%%%%%%%%%%%%%%%%%%%%%%%%%%%%%%%

\section{Introduction}

Several recent papers --- see %\cite{BCEZ,BEZ,BV,CFC}
[3]--[6], and references therein --- discussed the discrete
spectrum of the two-dimensional Pauli operator with a localized
magnetic field $B$, coming from an excess magnetic moment, $g>2$.
The most general result available concerns fields with a compact
support \cite{BCEZ}. In this situation the discrete spectrum is
nonempty whenever $B$ is nonzero, and its dimension is $1+[F]$
where $[F]$ is the integer part of the related flux (in natural
units).

The main aim of this letter is to extend this result to
non-compactly supported fields which satisfy a mild regularity
requirement and behave as $ \OO(|x|^{-2-\delta})$ for
$|x|\to\infty$. As long as we consider a powerlike bound, this is
an almost optimal condition, because $B$ has to be integrable. We
use a variational method to prove that if $B$ is a nontrivial
field with the stated decay and the absolute value of the flux
exceeds an integer $N$, then the Pauli operator with spin
antiparallel to the flux has at least $N+1$ bound states, counting
multiplicity. The variational proof follows the same idea as in
the compact-support case, but several modifications are needed.

Comparing to the mentioned theorem obtained in \cite{BCEZ} the
indicated result is slightly weaker giving one bound state less
for integer values of the flux. The reason is that without the
compact-support assumption we have less information about the
asymptotic behaviour of the Aharonov-Casher states used in the
construction, in particular, in case of integer flux the ``last"
one need not be bounded. On the other hand, we can replace the
sophisticated mollifier of \cite{BCEZ} by a simpler one.

The said difference is important in the case of zero flux when our
main result, Theorem~\ref{main thm} below, becomes trivial. It was
shown in \cite{BCEZ} that the existence of a discrete spectrum can
be then established for weak fields by the Birman-Schwinger
technique (see also \cite{BEZ} for the strong field case), and
moreover, that a bound state exists in this situation for both
spin orientations. The drawback of this result was that it
employed a (rather restrictive)  assumption about the decay of the
vector potential in the used gauge. We shall show that this
condition can be relaxed and the existence of weakly coupled bound
states can be proven under the mentioned assumptions on the
magnetic field alone.

%%%%%%%%%%%%%%%%%%%%%%%%%%%%%%%%%%%%%%%%%%%%%%%%%%%%%%%%%%%%%%%%%%%%%

\setcounter{equation}{0}
\section{Preliminaries}

We consider a two-dimensional electron interacting with a
non-homogeneous magnetic field $B= \partial_1A_2 -\partial_2A_1$
perpendicular to the plane. For the sake of simplicity, we employ
everywhere the natural units $2m= \hbar= c= e= 1$. The field
corresponds to a vector potential $A= (A_1,A_2)$ for which we
choose conventionally \cite{Th} the gauge $A_1= -\partial_2 \phi$,
$A_2=\partial_1 \phi$, where
\begin{equation} \label{phi}
\phi(x):= {1\over 2\pi} \int_{\R^2} B(y)\,\ln|x-y|\, d^2y \,,
\end{equation}
Below we give conditions under which the vector potential
components exist in the sense of distributions. The particle is
described by the Pauli Hamiltonian
\begin{equation} \label{Pauli}
H_P^{(\pm)}(A)=\left( -i\nabla-A(x)\right)^2\pm{g\over 2}\, B(x) =
D^*D+{1\over 2}\,(2\pm g) B(x)
\end{equation}
with $D:= (p_1\!-A_1)+ i(p_2\!-A_2)$, where the two signs
correspond to the two possible spin orientations. We are
particularly interested in the case when the electron has an
excess magnetic moment, $g>2$.

As in \cite{BCEZ} we shall suppose that $B\in L^1(\R^2)$. This
ensures the existence of a global quantity characterizing the
field,
\begin{equation} \label{flux}
F:= {1\over 2\pi}\, \int_{\R^2} B(x)\, d^2x\,,
\end{equation}
\ie, the total flux measured in the natural units $(2\pi)^{-1}$.
Without loss of generality we may assume $F\ge 0\,$; in that case
we will be interested primarily in the operator $H_P^{(-)}(A)$
which describes an electron with its magnetic moment parallel to
the flux.

The function (\ref{phi}) can be used to define the Aharonov-Casher
states which satisfy $D\chi_j = 0$ and thus yield zero-energy
solutions of the Pauli equation without the anomalous moment,
$g=2$. They are given by
\begin{equation} \label{AC ef's}
\chi_j(x)= e^{-\phi(x)}\, (x_1+ix_2)^j, \quad j=0,1,\dots
\end{equation}
For fields with a compact support we have $\chi_j(x)=
\OO(|x|^{-F+j})$ as $|x|\to\infty$ -- \cf \cite{AC},
\cite[Sec.7.2]{Th}. It means that if $F= N+\varepsilon$,
$\varepsilon\in (0,1]$ for a positive integer $N$, the operator
$H_P^{(-)}(A)$ with $g=2$ has $N$ zero energy eigenvalues.
Moreover, $\chi_{[F]}$ and possibly $\chi_{[F]-1}$ (in case that
$F$ is a positive integer; as usual, the symbol $[\cdot]$ denotes
the integer part) are zero energy resonances, since they solve the
equation $H_P^{(-)}(A)\chi_j=0$ and remain bounded at large
distances.

We shall assume the following:
\begin{description}
\item[(A.1)] $\:B(x) = \OO(|x|^{-2-\delta})$ for some $\delta>0$,
\item[(A.2)] $\:B \in L_{\mathrm{loc}}^{1+\epsilon}(\mathbb{R}^2)$
for some $\epsilon>0$.
\end{description}

\begin{remark}
If a positive number $\epsilon$ exists we can always choose it in
such a way that $\delta(1+\epsilon^{-1}) > 8$. Under the decay
requirement of (A.1) the second assumption means that $B \in
L^{1}(\mathbb{R}^2) \cap L^{1+\epsilon}(\mathbb{R}^2)$, in
particular, that the flux (\ref{flux}) makes sense and the same is
true for the integral (\ref{phi}) as we shall see in a while.
\end{remark}
The AC states now exist and their decay is given by the following
result.
\begin{proposition} \label{growth}
Assume (A.1) and (A.2). Then $\phi$ is locally bounded and to any
$\varepsilon>0$ there is a positive $R$ such that
\begin{equation} \label{bound}
\left| \phi(x) - F\ln|x| \right| <\varepsilon\, \ln|x|
\end{equation}
holds for all $|x|>R$.
\end{proposition}
{\em Proof:} Given a positive $c$ we denote $\langle y\rangle_c:=
\sqrt{c+y^2}$. Since $B\langle\cdot\rangle_c^{\delta/2} \in L^{1+
\epsilon}$, and $\langle\cdot\rangle_c^{-\delta/2} \ln|x-\cdot|
\in L^{1+ \epsilon^{-1}}$ for $\delta(1+\epsilon^{-1}) > 4$, the
H\"older inequality yields a bound on $|\phi(x)|$. To prove the
inequality (\ref{bound}), we denote $\BB_R:= \{x:\; |x| \le R \}$
and $\bar\BB_R:= \R^2\setminus\BB_R$. Furthermore, we set
\begin{equation} \label{F_0}
F_R:= {1\over 2\pi}\, \int_{\BB_R} B(x)\, d^2x\,,
\end{equation}
and
\begin{equation} \label{decomp}
\phi_R(x):= {1\over 2\pi} \int_{\BB_R} B(y)\,\ln|x-y|\, d^2y \,,
\quad \tilde\phi_R(x) := \phi(x) -\phi_R(x)\,.
\end{equation}
By assumption, to a given $\varepsilon>0$ there is $R_1$ such that
\begin{equation} \label{est1}
{1\over 2\pi} \int_{\bar\BB_{R_1}} |B(y)|\,\max\{1,\ln|y|\}\, d^2y
< {1\over 4}\,\varepsilon\,.
\end{equation}
It follows that
\begin{equation} \label{est2}
|F-F_{R_1}| < {1\over 4}\,\varepsilon\,.
\end{equation}
For any $R>0$ the quantity $F_R$ is the flux of a cut-off field
and $\phi_R$ is the corresponding ``potential". This allows us to
employ the above mentioned estimate \cite[Sec.7.2]{Th} by which
\begin{equation} \label{est3}
\phi_R(x) -F_R \ln|x| = \OO(|x|^{-1})
\end{equation}
as $|x|\to\infty$. Finally, we shall prove that
\begin{equation} \label{est4}
|\tilde\phi_{R_1}(x)| < {\varepsilon\over 4} \left( \ln|x| +
1+2\ln 2 \right) + c|x|^{-2-\delta}
\end{equation}
for some $c>0$ and all $|x|$ large enough. To this end we
decompose $\tilde\phi_{R_1} = \phi_1+\phi_2$ corresponding to the
integration over $|x-y|\le R_1$ and $|x-y|> R_1$, respectively.
The decay assumption yields
$$ |\phi_1(x)| \le {1\over 2\pi} \int_{|x-z|>R_1, |z|\le R_1}
c_1|x-z|^{-2-\delta} \ln|z|\, d^2z $$
for some $c_1>0$; we have used here the change of variable
$x-y=z$. We have $|x-z|^{-2-\delta} \le (|x|-R_1)^{-2-\delta} \le
|x|^{-2-\delta}$ for $|x|>R_1$, and therefore $\phi_1(x)
=\OO(|x|^{-2-\delta})$ as $|x|\to \infty$. Without loss of
generality we may suppose that $R_1>1$ and $|x|\ge 1$. Since
$|x-y| \le |x|+|y| \le (1+|x|)(1+|y|)$, the remaining part
$\phi_2(x)$ is then in view of
$$ 0 \le \ln|x-y| \le \ln(1+|x|) + \ln(1+|y|) \le 2\ln 2 + \ln|x|
+ \ln|y| $$
and of (\ref{est1}) estimated by the first term at \rhs of
(\ref{est4}). Putting now (\ref{est2})-(\ref{est4}) together we
find
$$ \left| \phi(x) - F\ln|x| \right| < {1\over 2}\,\varepsilon\,
\ln|x| + {\varepsilon\over 4} (1+2\ln 2) + c_2|x|^{-1}$$
with a suitable $c_2$. There is an $R_2$ such that the sum of the
last two terms is smaller than ${1\over 4}\,\varepsilon\, \ln|x|$
for $|x|>R_2$, so it is sufficient to set $R:= \max\{1,R_1,R_2\}$.
\quad \QED \vspace{2mm}

The above assumptions allow us to prove a stronger claim about the
regularity of $\phi$. Let us first recall two definitions
\cite{Ad}. Given an open ball $\BB^{(1)}$ centered at $x\in\R^n$
and an open ball $\BB^{(2)}$ not containing $x$, the set $C_x =
\BB^{(1)} \cap \{x+\lambda(y-x); y\in \BB^{(2)}, \lambda>0\}$ is
called a finite cone having vertex at $x$. An open domain
$\Omega\subset\R^n$ has the cone property if there exists a finite
cone $C$ such that each point $x\in\Omega$ is the vertex of a
finite cone $C_x$ contained in $\Omega$ and congruent to $C$. In
particular, every non-empty open ball in $\R^n$ has the cone
property. We shall employ the Sobolev imbedding theorem (\cf the
case C of the part~1 of Theorem~5.4 in \cite{Ad}) for the sets
\begin{eqnarray*}
 W^{m,p}(\Omega) &\!=\!& \{ u\in L^p(\Omega)\,:\:
    D^\alpha{u}\in L^p(\Omega)\mbox{ for }0\le|\alpha|\le m \}\,, \\
  C_{B}^{j}(\Omega) &\!=\!& \{ u\in C^j(\Omega)\,:\:
   D^\alpha{u}\,\mbox{ is bounded on }\,\Omega\mbox{ for }|\alpha|\le
   j\}\,.
\end{eqnarray*}
\begin{lemma}
  Let $\Omega$ be a domain in $\R^n$.  Suppose that $j$ and $m$
  are non-negative integers and $1\le{}p<\infty$, then the imbedding
  $W^{j+m,p}(\Omega)\to C_{B}^{j}(\Omega)$ exists provided $mp>n$.
\end{lemma}
\begin{lemma}
 {\rm (\cf \cite[Thm.~9.9]{GT})} Let $\Omega$ be a bounded domain
 in $\R^2$ and $f\in{}L^p(\Omega)$ with $1<p<\infty$. Define $w(x) =
 \int_\Omega \Gamma(x-y)f(y)\,dy$, where $\Gamma(x) =
 \frac{1}{2\pi}\ln|x|$; then $w\in W^{2,p}(\Omega)$.
\end{lemma}
Now we can state the indicated result:
\begin{proposition} \label{cont}
 Under the assumptions (A.1) and (A.2), $\phi$ is continuous in
 $\R^2$.
\end{proposition}
{\em Proof:} For arbitrary $x_0\in\R^2$ and $R>0$, we put
$\BB_R(x_0) = \{x\in \R^2; |x-x_0|<R\}$. We split $\phi$ as
follows:
$$
  \phi(x) =
  {1\over 2\pi} \int_{\BB_{2R}(x)} B(y)\,\ln|x-y|\, d^2y +
  {1\over 2\pi} \int_{\R^2\setminus \BB_{2R}(x)} B(y)\,\ln|x-y|\,
  d^2y.
$$
Since $\BB_{2R}(x)$ has the cone property, the first term at the
\rhs is in $W^{2,1+\epsilon}(\BB_{2R}(x))$, and thus also in
$C_B^0(\BB_{2R}(x))$ by the preceding two lemmas. On the other
hand, $\ln|x_1-y| - \ln|x_2-y| < \ln 3$ holds for any $x_1, x_2
\in \BB_R(x)$ and any $y \in \R^2\setminus \BB_{2R}(x)$, so
continuity of the second term follows by the Lebesgue
dominated-convergence theorem. \quad \QED \vspace{2mm}

\begin{remark}
Proposition~\ref{cont} can be proven in an alternative way. We
define a probability measure $\mu(dx)$ on $\mathbb{R}^2$ by
$$ \mu(dx) := \frac{1}{2\pi {\cal N}}\, |B(x)|\langle
x\rangle_{c}^{\delta/2}\, d^{2}x, $$
where ${\cal N} := \frac{1}{2\pi} \int_{\mathbb{R}^2}
|B(y)|\langle y\rangle_{c}^{\delta/2}d^{2}y$ is the normalization
factor, and a family of random variables $\left\{L_{x}\right\}_{x
\in  \BB_{\eta}(x_{0})}$ by
$$ L_{x}(y) := \langle y\rangle_c^{- \delta/2} \ln |x - y|, \qquad
y \in  \mathbb{R}^2,\,\, x \in \BB_{\eta}(x_{0})\,. \label{eq:rv}
$$
of which we can check that it is uniformly integrable, i.e.,
$$ \lim_{a\to\infty}\, \sup_{x \in \BB_{\eta}(x_{0})}
\int_{\left\{y\, |\, |L_{x}(y)| \geq a\right\}} |L_{x}(y)|\,
\mu(dy) = 0\,. $$
The argument leading to the last claim is based on simple
estimates but it is lengthy and we skip the details. The relation
$$ \lim_{x\to x_{0}}|\phi(x) - \phi(x_{0})| = {\cal N} \lim_{x\to
x_{0}} \int_{\mathbb{R}^2} |L_{x}(y) - L_{x_{0}}(y)|\mu(dy) = 0,,
$$
then follows from the abstract result given in \cite[Theorem
3.7.4]{Ito} or \cite[Prop.~II.5.4]{Ne}.
\end{remark}

We will also need a bound on the vector potential, or
equivalently, on the gradient of the potential (\ref{phi}). Its
components are given by
\begin{equation} \label{grad}
(\partial_i\phi)(x)= {1\over 2\pi} \int_{\R^2}
B(x-z)\,{z_i\over|z|^2}\, d^2z \,,
\end{equation}
at least for large enough $|x|$ where $B$ is bounded. While in
general they behave as $\OO(|x|^{-1})$, in case of zero flux we
have a stronger result.
\begin{proposition} \label{grad decay}
In addition to the stated integrability and decay assumptions,
suppose that $\int_{\R^2} B(y)\, d^2y =0$; then there is $\mu>0$
such that $(\nabla\phi)(x)= \OO(|x|^{-1-\mu})$ as $|x|\to\infty$.
\end{proposition}
{\em Proof:} Consider $(\partial_1\phi)(x)$; the argument for the
other component is similar. We write it as $\sum_{j=1}^4
A_2^{(j)}(x)$, where the different contributions correspond to
integration over the regions where $|x-z|$ and $|z|$ are
respectively smaller and greater that $R_3$. The last named number
depends on $|x|$ and will be specified later.

Since $|z|\ge ||x|-|x-z||$, the term $A_2^{(1)}(x)$ with $|x-z|\le
R_3$ and $|z|\le R_3$ is zero provided
\begin{equation} \label{Rcond1}
|x|>2R_3\,.
\end{equation}
The term $A_2^{(2)}(x)$ obtained by changing the first inequality
to $|x-z|> R_3$ is estimated easily as
\begin{equation} \label{Rcond2}
\left|A_2^{(2)}(x)\right| \le {c_1\over 2\pi}\, R_3^{-2-\delta}
\int_{|z|\le R_3} {d^2z\over |z|} = c_1 R_3^{-1-\delta}\,.
\end{equation}
The third term corresponding to integration over $M_3:= \{z:\:
|x-z|\le R_3,\, |z|> R_3 \}$ is the most complicated. Combining
the decay and the zero-flux assumptions we get
\begin{equation} \label{approx zero}
\left|\int_{|y|\le R_3} B(y)\, d^2y\right| = \left|\int_{|y|> R_3}
B(y)\, d^2y\right| \le {2\pi c_1\over \delta}\, R_3^{-\delta}\,.
\end{equation}
Next we split the field into the positive and negative part,
$B=B_+-B_-$, and write
$$ A_2^{(3)}(x) = {1\over 2\pi} \int_{M_3}
B_+(x-z)\,{z_i\over|z|^2}\, d^2z - {1\over 2\pi} \int_{M_3}
B_-(x-z)\,{z_i\over|z|^2}\, d^2z\,. $$
It is straightforward to check that $\left|z_1|z|^{-2} -
|x|^{-1}\cos\theta \right| \le 5R_3|x|^{-2}$ holds for
$R_3|x|^{-1}$ small enough, where $\theta$ is the angle
corresponding to $x$ in polar coordinates. We use this inequality
to get an upper and lower bound to $z_i|z|^{-2}$ in the above
integrals. Then we add and subtract ${\cos\theta + 5R_3|x|^{-1}
\over 2\pi|x|} \int_{M_3} B_-(x-z)\, d^2z$ obtaining thus
\begin{eqnarray} \label{Rcond3}
A_2^{(3)}(x) &\!\le\!& {\cos\theta + 5R_3|x|^{-1} \over 2\pi|x|}
\int_{M_3} B(x-z)\, d^2z + {5R_3\over \pi|x|^2} \int_{M_3}
B_-(x-z)\, d^2z \nonumber \\ &\!\le\!& {2c_1\over\delta}\,
R_3^{-\delta}\, {\cos\theta\over|x|} + {c_3\over |x|^2}\, R_3
\end{eqnarray}
and an analogous lower bound, where in the second step we have
used (\ref{approx zero}) and the integrability of $B$. If we
choose $R_3= |x|^{1-\eta}$ for $\eta<1$ we get
\begin{equation} \label{Rcond4}
\left|A_2^{(3)}(x)\right| \le c_4 \max\{|x|^{-1-\delta(1-\eta)},
|x|^{-1-\eta}\}
\end{equation}
for some $c_4>0$ and large $|x|$; at the same time the condition
(\ref{Rcond1}) will be satisfied. The remaining term with
$|x-z|>R_3$ and $|z|>R_3$ is estimated by
$$ \left|A_2^{(4)}(x)\right| \le {1\over 2\pi R_3}
\int_{|x-z|>R_3} B(x-z)\,d^2z \le {c_1\over R_3}\,
\int_{R_3}^{\infty} r^{-1-\delta} dr = {c_1\over\delta}\,
R_3^{-1-\delta}\,. $$
With our choice, $R_3= |x|^{1-\eta}$, we get from here and
(\ref{Rcond2})
\begin{equation} \label{Rcond5}
\max\left\{\left|A_2^{(2)}(x)\right|, \left|A_2^{(4)}(x)\right|
\right\} \le c_5|x|^{-1-\delta+\eta(1+\delta)}\,,
\end{equation}
so it is sufficient to set $\eta< \delta(1+\delta)^{-1}$ to get a
decay power smaller than $-1$. $\;\QED$

%%%%%%%%%%%%%%%%%%%%%%%%%%%%%%%%%%%%%%%%%%%%%%%%%%%%%%%%%%%%%%%%%%%%%%%

\setcounter{equation}{0}
\section{The main result}

Now we are ready to extend the result of \cite{BCEZ} about the
existence and number of bound states to fields without a compact
support.

\begin{theorem} \label{main thm}
Let $B$ be nonzero, satisfying (A.1) and (A.2), and let the
corresponding flux be $F=N+\eta$ for some $N\in\N_0$ and $\eta>0$.
Then the operator $H_P^{(-)}(A)$ has for $g>2$ at least $N+1$
isolated eigenvalues in $(-\infty,0)$, multiplicity being counted.
\end{theorem}
\noindent {\em Proof:} First we need to know that the essential
spectrum covers the positive halfline. Since the last term in
(\ref{Pauli}) can be viewed as a potential which is
$\Delta$-compact, it follows from \cite[Thm.~6.1]{CFKS} and
\cite[Sec.~XIII.4]{RS} that
\begin{equation} \label{ess}
\sigma_\mathrm{ess}(H_P^{(\pm)}(A))= [0,\infty)\,.
\end{equation}
In view of the minimax principle, it is then sufficient to find an
$(N+1)$-dimensional subspace in $L^2(\R^2)$ on which the quadratic
form
$$ \psi\mapsto  (\psi,H_P^{(-)}(A)\psi)= \int_{\R^2}
|(D\psi)(x)|^2\, d^2x -{1\over 2}\,(g-2) \int_{\R^2} B(x)
|\psi(x)|^2\, d^2x $$
is negative. We will employ trial functions $\psi_{\alpha}$ of the
following form
\begin{equation} \label{psi}
\psi_{\alpha}(x) = \sum_{j=0}^{N} \alpha_j \left(
f_{\varrho}(r)\chi_j(x) +\varepsilon h_j(x) \right)
\end{equation}
with $\alpha\in\C^{N+1}$; it is clearly sufficient to consider the
unit sphere, $|\alpha|=1$. Here $f_{\varrho}$ is a mollifier which
will be chosen as $f_{\varrho}(x):= f(|x|/\varrho)$ for a
real-valued function $f\in C_0^{\infty}(\R_+)$ such that $f(u)=1$
for $u\le 1$ and $f(u)=0$ for $u\ge 2$. The functions  $h_j\in
C_0^{\infty}(\BB_{\varrho})$ will be specified later. By a direct
computation,
\begin{eqnarray}\label{energy form}
(\psi_{\alpha},(D^*D+\mu B)\psi_{\alpha}) &\!=\!& \sum_{j,k=0}^{N}
\bar\alpha_j \alpha_k \left\{ \int_{\bar\BB_{\varrho}}
\left|f'_{\varrho}(r)\right|^2 (\bar\chi_j \chi_k)(x)\, d^2x
\right. \\ &\!+\!& \varepsilon^2\, \int_{\BB_{\varrho}} (D\bar
h_j)(x) (Dh_j)(x)\, d^2x + \mu\left[ \int_{\R^2} (f_{\varrho}^2 B
\bar\chi_j \chi_k)(x)\,d^2x \right. \nonumber \\ &\!+\!&
\varepsilon\, \int_{\BB_{\varrho}} ((\bar h_j\chi_k + \bar\chi_j
h_k)B)(x)\, d^2x + \left. \left. \varepsilon^2
\int_{\BB_{\varrho}} (B \bar h_j h_k)(x)\, d^2x \right] \right\}
\nonumber
\end{eqnarray}
where we have employed $D\chi_j=0$ together with  the fact that
$h_j$ and $f'_{\varrho}$ have by construction disjoint supports:
$D \Sigma_j \alpha_j f_{\varrho}\chi_j = 0$ holds inside
$\BB_{\varrho}$ so $D\psi_{\alpha} = \varepsilon \Sigma_j \alpha_j
Dh_j$ there, while outside we have instead $D\psi_{\alpha} =
Df_{\varrho}\Sigma_j \alpha_j \chi_j = \psi_{\alpha} (-ix_1+x_2)
|x|^{-1} f'_{\varrho}$. We have to show that the \rhs is negative
as long as $\mu<0$, in particular, for $\mu= -{1\over
2}\,(g\!-\!2)$.

The mollifier is necessary since the sum (\ref{psi}) contains in
general terms which are not $L^2$. The corresponding contribution
to the energy form, \ie, the first term at the \rhs of
(\ref{energy form}) is positive and we have to make it small.
Since $f'_{\varrho}$ is supported in $\BB_{2\varrho}$, it follows
from Proposition~\ref{growth} that
\begin{equation} \label{tail bound}
 {1\over \varrho^2} \int_{\bar\BB_{\varrho}}
\left|f'\left(|x|\over \varrho \right) \right|^2\,  \left|
\sum_{j=0}^{N} \alpha_j \chi_j(x) \right|^2\, d^2x \le
{4\pi\|f'\|^2_{\infty}\over 1+\varepsilon+N-F}
(2\varrho)^{2(N-F+\varepsilon)}
\end{equation}
provided $\varrho>R$. Without loss of generality we may assume
$\eta\in (0,1]$. Choosing then $\varepsilon\in (0,\eta)$, we
obtain a bound which tends to zero as $\varrho\to \infty$, and
therefore it allows us to handle the trial function tails.

The main part of the argument consists of checking that there
exists a positive constant $\beta$ such that
\begin{equation}
  \label{eq:positive}
  \int_{\R^2} B(x)\left|f_{\varrho}(x) \sum_j
  \alpha_j\chi_j(x)\right|^2\, d^2x\, > \beta
\end{equation}
holds for $\varrho$ large enough and any $\alpha$. We shall do it
by {\em reductio ad absurdum} assuming the opposite. Now we have
to specify the functions $h_j$. We set $h_j:= h\chi_j$ for a
real-valued $h\in C_0^{\infty}(\R_+)$, in which case the next term
linear in $\varepsilon$ acquires the form
$$ 2\varepsilon\, \int_{\R^2} \left| \sum_{j=0}^{N} \alpha_j
\chi_j(x) \right|^2 h(x) B(x)\, d^2x\,. $$
Since $B$ is nonzero by assumption, and $\sum_j \alpha_j\chi_j$ is
a product of a positive function $e^{-\phi}$ and a polynomial
having thus at most isolated zeros, one can choose $h$ in such a
way that the last expression is negative for any $\alpha$.
Moreover, as a continuous function of $\alpha$ on the surface of a
hypersphere it reaches a minimum there which is also negative.
This implies that the sum of the second, fourth and fifth terms of
Eq.~(\ref{energy form}), denoted as $S$, tends to \(0\) from below
as $\varepsilon$ tends to $0$. Hence there is a number $\beta
> 0$ such that for $\varrho$ large enough and any $\alpha$,
one can find $h=h_{\alpha,\varrho}$ and $\varepsilon_{\alpha,
\varrho}$ for which $S=-2\mu\beta$ holds. Suppose that
\begin{equation}
  \int_{\R^2}B(x)\left|f_{\varrho}(x) \sum_j
  \alpha_j\chi_j(x)\right|^2\,d^2x \le \beta
\end{equation}
holds true. Choosing then $h_{\alpha,\varrho}$ and
$\varepsilon_{\alpha, \varrho}$ in the described way, we get
\begin{equation}
  S + \mbox{the third term of Eq.~(\ref{energy form})}
  \le -2\mu\beta + \mu\beta = -\mu\beta < 0\,.
\end{equation}
However, in view of (\ref{tail bound}) we have
\begin{equation}
  \mbox{the first term of Eq.~(\ref{energy form})}
  \le
  {4\pi\|f'\|^2_{\infty}\over 1+\varepsilon-\eta}
  (2\varrho)^{-2(\eta-\varepsilon)}
  \to 0
\end{equation}
as \(\varrho\) tends to \(\infty\), so the \rhs of (\ref{energy
form}) is negative for \(\varrho\) large enough. The argument can
be carried over for any fixed value of $\mu$, in particular, for
$\mu=2$. In that case, however, the supersymmetry property,
$D^*D+2B=DD^*$, applied to the \lhs of (\ref{energy form}) leads
to the absurd conclusion $\|D^*\psi_{\alpha}\|^2 < 0$, proving
thus Eq.~(\ref{eq:positive}).

This means that the trial functions can be finally chosen in the
form (\ref{psi}) with $\varepsilon=0$. The energy form is then
estimated by
\begin{equation} \label{eform bound}
(\psi_{\alpha},H_P^{(-)}(A)\psi_{\alpha})<
{4\pi\|f'\|^2_{\infty}\over 1+\varepsilon-\eta}
(2\varrho)^{-2(\eta-\varepsilon)}- {1\over 2}\,(g-2) \int_{\R^2}
B(x)|\psi_{\alpha}(x)|^2\, d^2x\,,
\end{equation}
where the second term at the \rhs is smaller that $-{1\over
2}(g-2)\beta$ and dominates for $\varrho$ large enough. With our
choice of the mollifier, $\psi_{\alpha}$ is within $\BB_{\varrho}$
just a linear combination of the Aharonov-Casher states (\ref{AC
ef's}). Since the latter are easily seen to be linearly
independent we have accomplished the task of construction the
sought $(N\!+\!1)$-dimensional subspace. $\:\QED$

%%%%%%%%%%%%%%%%%%%%%%%%%%%%%%%%%%%%%%%%%%%%%%%%%%%%%%%%%%%%%%%%%%%%%

\setcounter{equation}{0}
\section{Zero flux case}

In distinction to the analogous result in \cite{BCEZ},
Theorem~\ref{main thm} says nothing about the situation when
$F=0$. For radially symmetric strong and weak fields the bound
state existence is established in \cite{BEZ} and \cite{BCEZ},
respectively. For weak  fields without the rotational symmetry we
can employ the method of Sec.~6 in \cite{BCEZ}, but without the
assumption about the decay of $\nabla\phi$ used there. We need
only a slightly stronger regularity requirement:
\begin{description}
\item[(A.2')] $\:B \in L_{\mathrm{loc}}^2(\mathbb{R}^2)$.
\end{description}
Recall that the said idea in \cite{BCEZ} is based on the
weak-coupling behaviour of two-dimensional Schr\"odinger operators
with a potential depending on a coupling constant in a nonlinear
way, specifically
\begin{equation} \label{2D Hamiltonian}
H(\lambda) = -\Delta + \lambda V_1(x) + \lambda^2 V_2(x)
\end{equation}
with $V_j \in L^{1+\delta}(\R^2) \cap L(\R^2,
(1+|x|^{\delta})\,d^2x),\, j= 1,2$.
\begin{lemma} \label{weak bd}
{\rm \cite[Sec.~4]{BCEZ}} Suppose that $\int V_1(x)\,d^2x = 0$ and
define
\begin{equation} \label{c_2}
\gamma_2 \equiv \gamma_2(V_1,V_2) := {1\over 2\pi}\int
V_2(x)\,d^2x + {1\over 4\pi^2}\int V_1(x)\, \ln|x-y|\, V_1(y)\,
d^2x\, d^2y\,.
\end{equation}
The operator (\ref{2D Hamiltonian}) has a weakly bound state for
small nonzero $\lambda$ iff the quantity (\ref{c_2}) is negative.
In that case the eigenvalue is $\epsilon(\lambda)=
-e^{2/u(\lambda)}$ with $u(\lambda)= \gamma_2\lambda^2+
\OO(\lambda^3)$.
\end{lemma}
\begin{lemma}
Under (A.1) and (A.2') the function $\phi\in W^{1,2}(\R^2)$.
\end{lemma}
\noindent {\em Proof:} The function $\phi = {1\over 2\pi}\,
B\ast\ln|\cdot|$ belongs to the first Sobolev space if the
integral $\int (1+|k|^2) |\hat B(k)|^2 |k|^{-2} d^2k$ is finite.
The assumptions imply $B\in L^2$, and therefore also $\hat B\in
L^2$; hence we have to check only its convergence around $k=0$. We
have $\hat B(0)=F=0$, so
$$ \hat B(k) = {1\over 2\pi} \int_{\R^2} B(x) \left( e^{ikx}-1
\right) d^2x\,. $$
Further we decompose $\R^2= \BB_R \cup \tilde\BB_R$ as in the
proof of Proposition~\ref{growth} with the circular boundary
situated in the region where $B$ is bounded. We estimate
$|e^{ikx}-1|$ by $kR$ in the inner region and by $2|kx|^{\eta}$
with $\eta\in (0,1)$ outside obtaining
$$ |\hat B(k)| \le c_1|k| + {1\over\pi}\, |k|^{\eta}
\int_{\tilde\BB_R} |x|^{\eta} |B(x)|\, d^2x $$
for some $c_1>0$. Choosing now $\eta$ sufficiently small we can
make the last integral finite; this yields $|\hat B(k)|^2=
\OO(k^{2\eta})$ around the origin. \quad \QED \vspace{2mm}

Now we can prove the following result.
\begin{theorem} \label{zero flux thm}
Let a nonzero $B$ with $F=0$ satisfy (A.1) and (A.2'). Then each
of the operators $H_P^{(\pm)}(\lambda A)$ with $g>2$ has for small
nonzero $\lambda$ a bound state whose energy satisfies the bound
\begin{equation} \label{weak bound}
\epsilon^{(\pm)}(\lambda) < -\, \exp\left\{ -\left(
{c\lambda^2\over 16\pi} (g^2\!-4)\, \int_{\R^2} A(x)^2\, d^2x
\right)^{-1} \right\}
\end{equation}
for any fixed $c\in (0,1)$ and $\lambda$ small enough.
\end{theorem}
\noindent {\em Proof:} It is established in \cite{BCEZ} that the
gradient term $2iA\cdot\nabla$ does not contribute to the energy
form for real-valued functions, and therefore $H_P^{(\pm)}(\lambda
A)$ can be estimated from above by the operators (\ref{2D
Hamiltonian}) with
\begin{equation} \label{potential coefficients}
V_1(x)= \,\pm\,{g\over 2}\,B(x)\,, \qquad V_2(x)= A(x)^2\,.
\end{equation}
It remains to evaluate the coefficient (\ref{c_2}). Since $|A|$ is
square integrable by the preceding lemma and  $|A(x)| =
|(\nabla\phi)(x)|$, the first Green identity together with the
equation $\Delta\phi=B$ and the Gauss theorem yield
\begin{equation} \label{identity}
\int_{\R^2} A(x)^2 d^2 x =  \lim_{R\to\infty} \oint_{\partial
\BB_R} \phi(x)(\nabla\phi)(x)\cdot d\vec\sigma(x) -
\lim_{R\to\infty} \int_{\BB_R} \phi(x)B(x)\, d^2x\,.
\end{equation}
Substituting from (\ref{phi}) to the last term we see that it
remains to establish that the first term at the \rhs vanishes as
$R\to\infty$. However, this follows readily from
Propositions~\ref{growth} and \ref{grad decay}. $\;\QED$

%%%%%%%%%%%%%%%%%%%%%%%%%%%%%%%%%%%%%%%%%%%%%%%%%%%%%%%%%%%%%%%%%%%%%%%%%%%%

\subsection*{Acknowledgments}

We thank the referee for a useful comment. The research has been
partially supported by GAAS and Czech Ministry of Education under
the contracts 1048801 and ME170. M.H. was supported by
Grant-in-Aid 11740109 for Encouragement of Young Scientists from
Japan Society for the Promotion of Science.


\begin{thebibliography}{99}
%
\bibitem{Ad}
R.A.~Adams: {\em Sobolev Spaces}, Academic Press, New York 1975.
\vspace{-1.8ex}
%
\bibitem{AC}
Y.~Aharonov, A.~Casher: Ground state of a spin--1/2 charged
particle in a two--dimensional magnetic field, {\em Phys. Rev.}
{\bf A19} (1979), 2641--2642. \vspace{-1.8ex}
%
\bibitem{BCEZ}
F.~Bentosela, R.M.~Cavalcanti, P.~Exner, V.A.~Zagrebnov: Anomalous
electron trapping by localized magnetic fields, {\em J. Phys.}
{\bf A32} (1999), 3029--3039. \vspace{-1.8ex}
%
\bibitem{BEZ}
F.~Bentosela, P.~Exner, V.A.~Zagrebnov: Electron trapping by a
current vortex, {\em J. Phys.} {\bf A31} (1998), L305--311.
\vspace{-1.8ex}
%
\bibitem{BV}
M.~Bordag, S.~Voropaev: Charged particle with magnetic moment in
the Aharonov-Bohm potential, {\em J. Phys.} {\bf A26} (1993),
7637--7649. \vspace{-1.8ex}
%
\bibitem{CFC}
R.M.~Cavalcanti, E.S.~Fraga, C.A.A.~de Carvalho: Electron
localization by a magnetic vortex, {\em Phys. Rev.} {\bf B56}
(1997), 9243--9246. \vspace{-1.8ex}
%
\bibitem{CFKS} H.L.~Cycon, R.G.~Froese, W.~Kirsch, B.~Simon:
{\em Schr\"odinger Operators with Applications to Quantum
Mechanics and Global Geometry,} Springer, Berlin 1987.
\vspace{-1.8ex}
%
\bibitem{GT}
D.~Gilbarg, N.S.~Trudinger: {\em Elliptic Partial Differential
Equations of Second Order}, Springer-Verlag, Berlin 1983.
\vspace{-1.8ex}
%
\bibitem{Ito} K.~Ito: {\em Introduction to Probability Theory,}
Cambridge University Press 1984. \vspace{-1.8ex}
%
\bibitem{Ne} J.~Neveu: {\em Mathematical Foundations of the Calculus
of Probability,} Holden-Day, San Francisco 1965. \vspace{-1.8ex}
%
\bibitem{RS} M.~Reed, B.~Simon: {\em Methods of Modern Mathematical
Physics, IV. Analysis of Operators,} Academic Press, New York
1978.\vspace{-1.8ex}
%
\bibitem{Th}
B.~Thaller: {\em The Dirac equation}, Springer, Berlin 1992.
\vspace{-1.8ex}
%
   \end{thebibliography}
\end{document}